\documentclass[journal]{an}

\usepackage{caption}
      \usepackage{graphicx}

\usepackage{xcolor}

\begin{document}

\Pagespan{1}{}
\Yearpublication{2018}%
\Yearsubmission{2017}%
\Month{9}%
\Volume{999}%
\Issue{0}%
\DOI{asna.201400000}%

\title{Period changes in the RR Lyrae stars of NGC 6171 (M107)}

\author{A. Arellano Ferro\inst{1}, P. Rosenzweig\inst{2}, A. Luna\inst{1}, D.
Deras\inst{1}, S. Muneer\inst{3}, Sunetra Giridhar\inst{3}, R. Michel\inst{4}}

\titlerunning{RR Lyrae period changes in NGC 6171}
\authorrunning{Arellano Ferro et al.}

\institute{
Instituto de Astronom\'ia, Universidad Nacional Aut\'onoma de M\'exico.
Ciudad Universitaria, CP 04510, M\'exico: (armando@astro.unam.mx)
\and 
Grupo de Astrof\'isica Te\'orica, Facultad de Ciencias, Universidad de Los Andes,
Venezuela.
\and
Indian Institute of Astrophysics, Koramangala, 560034, Bangalore, India
\and
Observatorio Astron\'omico Nacional, Instituto de Astronom\'ia Universidad Nacional
Aut\'onoma de M\'exico, Ap. P. 877, Ensenada, BC 22860, M\'exico
}

\received{2017}
\accepted{12-01-2018}
\publonline{XXXX}


\abstract{%
Based on photometric data obtained between 1935 and 2017, $O-C$ diagrams were
built for 22
RR Lyrae stars in the globular cluster NGC 6171, leading to the discovery of secular
period changes in 4 variables for which we have calculated their period change rates
$\beta$. In contrast we find that $82\%$ of the sample stars have stable periods over
the last
82 years. For the stable period stars, the whole data base has been employed to refine
their periods. Among the period changing stars, three (V10, V12 and V16) have
decreasing periods larger than expected from stellar evolution. Despite these
individual cases of significant period change rate, the golbal average of the
measured period changes in the cluster is basically zero, in consonance with
theoretical predictions for clusters with redder horizontal branches. 
The hitherto unpublished observations, now brought into
public domain, are employed to calculate a set of times of maximum light which are
used in the present analysis.}

\maketitle

\section{Introduction}

The study of secular period changes of RR Lyrae stars (RRLs) in globular clusters,
may play a decisive role in testing horizontal branch (HB) evolution models. However,
measuring secular period changes from observations
is difficult since accurate observations over a very long time-base
are required. Only a few clusters have been studied from data covering more 
than 60 years, e.g.; M3 (Corwin \& Carney 2001, Jurcsik et al. 2012), M5 (Szeidl et
al. 2011; Arellano Ferro et al. 2016), NGC 6934 (Stagg \& Wehlau 1980), 
M14 (Wehlau \& Froelich 1994), M15 (Silbermann \& Smith 1995), NGC 7006 (Wehlau,
Slawson \& Nemec 1999), and  $\omega$ Cen (Jurcsik et al. 2001). Theory predicts that
blueward or redward evolution near the zero-age
horizontal branch (ZAHB) is slow and produces very small period change rates, except
towards the end of the HB evolution, when the values of $\beta=\dot{P}$ can be
between $+0.1$ and $+0.15$ d~Myr$^{-1}$ (Lee 1991). However, in several of the above
studies, 
stars with significantly large values of $\beta$, both positive and negative, have
been reported. These high values may be the result
of non-evolutionary effects, like stochastic processes related to mixing events 
in the core of the star (Balazs-Detre \& Detre 1965, Sweigart \& Renzini 1979), or
to the fast crossing of the instability strip of pre-ZAHB stars
on their evolution to the  blue, with $\beta \leq -0.3$ d~Myr$^{-1}$ (Silva-Aguirre et
al. 2008).

In the present investigation we have focused our attention on the globular
cluster 
NGC 6171 (M107 or C1629-129 in the IAU nomenclature, $\alpha = 16^{\mbox{\scriptsize
h}}
32^{\mbox{\scriptsize m}} 31.86^{\mbox{\scriptsize s}}$, $\delta = +
13\degr 03\arcmin
13.6\arcsec$  J2000,  galactic coordinates $l = 3\degr.37$, $b =
+23\degr.01$).

A previous study of secular period changes in NGC 6171 was published by Coutts \&
Sawyer Hogg (1971), whom included photographic data of 22 RRLs taken between 1935 and
1970, however, the times of maximum light were rather scanty. A commentary
of the period change rates based on statistical theoretical grounds was published by
Gryzunova (1979a,b). Suplemented with data
from Las Campanas 1972-1991 (now published in the present work); Las
Campanas 1993-1994 (Clement \& Shelton 1997) and our Hanle
2015-2016 and San Pedro M\'artir 2017 CCD observations, the time-base
extends to $\sim$82 years, constitutes a significant improvement and encourages
a new approach to the study of the secular period changes of the RRLs of this cluster.

The photometric data were used to calculate as many times of maximum light as possible
and these were employed to investigate the
secular period behavior of 22 RRLs in NGC 6171. The sources and temporal
distribution of the data are indicated in Table \ref{archDATA}. 

The present paper is organized as follows; in $\S$ \ref{OBS} we briefly describe 
our 2015-2017 observations, in $\S$ \ref{Previous} we summarise the 1935-1994 data 
taken from the literature, in $\S$ \ref{OmC} the $O-C$ method is described. 
In $\S$ \ref{Tmax} the approach used to estimate
the times of maximum brightness is explained, and it contains the individual
$O-C$ diagrams and the resulting refined periods and period change rates. In $\S$
\ref{EVOL_HB} we discuss our results in the context of stellar evolution; and
finally, in $\S$ \ref{summary} our conclusions are summarised.

\begin{table}
\caption{Sources of photometric data of the RRLs in NGC 6171.}
\centering
\begin{tabular}{lcccc}
\hline
Authors  &  years & band  \\
\hline
Oosterhoff (1938)&1935&$pg$ \\
Coutts \& Sawyer Hogg (1971) &1946-1970 &$pg$\\
Kukarkin (1961)&1959-1960&$pg$\\
Mannino (1961)&1959-1960&$pg$\\
Dickens (1970)&1966-1967 &$B_{pg}$ and $V_{pg}$\\
Table 3 (this paper) & 1972-1991& $B_{pg}$ \\
Clement \& Shelton (1997)&1993-1994 & CCD $V$\\
Table 2 (this paper) & 2015-2017& CCD $V$ \\
\hline
\end{tabular}
\label{archDATA}
\end{table}

\begin{table}
\caption{2015-2017 $VI$ photometry data of the RRLs in NGC 6171. A full version
of this table is available in electronic format (see Supporting Information).}
\centering
\begin{tabular}{ccccc}
\hline
Variable &Filter & HJD & $M_{\mbox{\scriptsize std}}$ &$\sigma_{m}$\\
Star ID  &    & (d) & (mag)     & (mag)  \\
\hline

 V4&  V& 2457200.14416&  15.673&  0.006\\ 
 V4&  V& 2457200.14769&  15.667&  0.006\\
\vdots   &  \vdots  & \vdots & \vdots &  \vdots\\
 V4&  I& 2457200.13190&  14.797&   0.009\\ 
 V4&  I& 2457200.13824&  14.787&   0.009\\
\vdots   &  \vdots  & \vdots & \vdots  & \vdots\\
 V5&  V& 2457200.14416&  15.287&  0.004\\ 
 V5&  V& 2457200.14769&  15.286&  0.004 \\
\vdots   &  \vdots  & \vdots & \vdots & \vdots\\
 V5&  I& 2457200.13190&  14.327&  0.007\\ 
 V5&  I& 2457200.13824&  14.302&  0.007\\
\vdots   &  \vdots  & \vdots & \vdots &  \vdots\\
\hline
\end{tabular}
\label{PHOT15-17}
\end{table}

\section{Observations}
\label {OBS}

The most recent $V$ CCD time-series observations used in this paper substantially
extend the
time baseline, in many cases to $\sim$ 82 years. These observations were performed on
4 nights, between June 26, 2015 to May 5, 2016,
 with the 2.0~m telescope at the Indian Astronomical Observatory (IAO), Hanle,
India. For 7 nights between June 29 to July 5, 2017, data were obtained with the
0.84~m telescope at the San Pedro M\'artir Observatory (SPM), M\'exico.
A total of 292 images obtained in the Johnson-Kron-Cousins $V$ filter, are used
for the purpose of the present analysis. $I$ images were also obtained. The
instrumental system was converted into the Landolt-Johnson/Kron-Cousins standard
system via standard stars in the field of the cluster provided by Stetson (2000). The
standard system $VI$ magnitudes and their uncertainties for the RRLs are published in
electronic format and we present a small portion of it as Table \ref{PHOT15-17}.

\begin{table*}
\caption{Las Campanas photographic data (B mag) taken between 1972 and 1991 by C.
Clement and previously unpublished. This is an extract
from the full table, which is only available with the electronic version of the
article (see Supporting Information).}
\label{unpub}
\centering
\begin{tabular}{cccccccccccc}
\hline
  HJD &    V2 &  V3  & V4  & V5 &  V6  & V7  & V8  & V9 &  V10 & V11 & V12\\
-2~400~000 & &&&&&&&&&\\
\hline
41446.819 &  16.86&16.44&15.56&15.91&16.02&15.33&16.44&15.77&16.84&16.07&16.89\\
41447.640 &   -- & -- &  --   & --  & -- &  -- & --  & --  & --  & -- &  --\\
41447.676 &  15.84&15.89&15.56&16.29&15.60&16.29&15.25 & -- &16.86&16.61&16.61\\
. . . &&&&&&&&&&&\\
\hline
  HJD &   V13 &  V14 &  V15  & V16  & V17 &  V18  & V19  & V20 & V21 & V23 & V24\\
-2~400~000 & &&&&&&&&&&\\
\hline
41446.819 & 16.79&16.89&15.66&16.84&16.66&15.24&15.91&16.70&16.95&16.15& --  \\
41447.640 & --   & --  &15.77&16.05&15.77&15.99&16.44&  -- &  -- &15.89& --  \\ 
41447.676 & 16.98&16.73&15.79&16.42&15.93&16.02&15.89&16.02&17.16&15.73&16.29\\
. . . &&&&&&&&&&&\\
\hline
\end{tabular}
\end{table*}

\subsection{The 1935 - 1994 data of NGC~6171}
\label{Previous}

The data taken between 1935 and 1994 have been systematically assembled by Prof.
Christine M. Clement,
who kindly made them available to us.
The original sources are summarised in Table \ref{archDATA}. The published
data have been taken as given in the original papers without any further
manipulation. The observations from the years 1972-1991 were obtained by C.
Clement
      with the University of Toronto 61-cm telescope at the Las Campanas 
      Observatory of the Carnegie Institution of Washington.  
      A total of 420 photographs were taken on plates with 103aO 
      emulsion, exposed through a GG385 filter. The plates were measured 
      on a Cuffey iris photometer.  Some of these data were used, but not 
      published, in the study of the Fourier parameter $\phi_{31}$, by 
      Clement et al. (1992).
We are now publishing these data in an electronic format and a small fraction of the
table is
included in the printed version of this paper as Table \ref{unpub}. These
data were taken in an almost yearly basis and the light curves of most variables
are well covered, allowing a good estimation of the time of maximum light
nearly every year. Thus, this set of data is crucial for the interpretation of the
$O-C$ diagrams.

\subsection{The 2015 - 2016 data of NGC~6171}
\label{Present}

\begin{figure*}
\begin{center}
\includegraphics[scale=0.95]{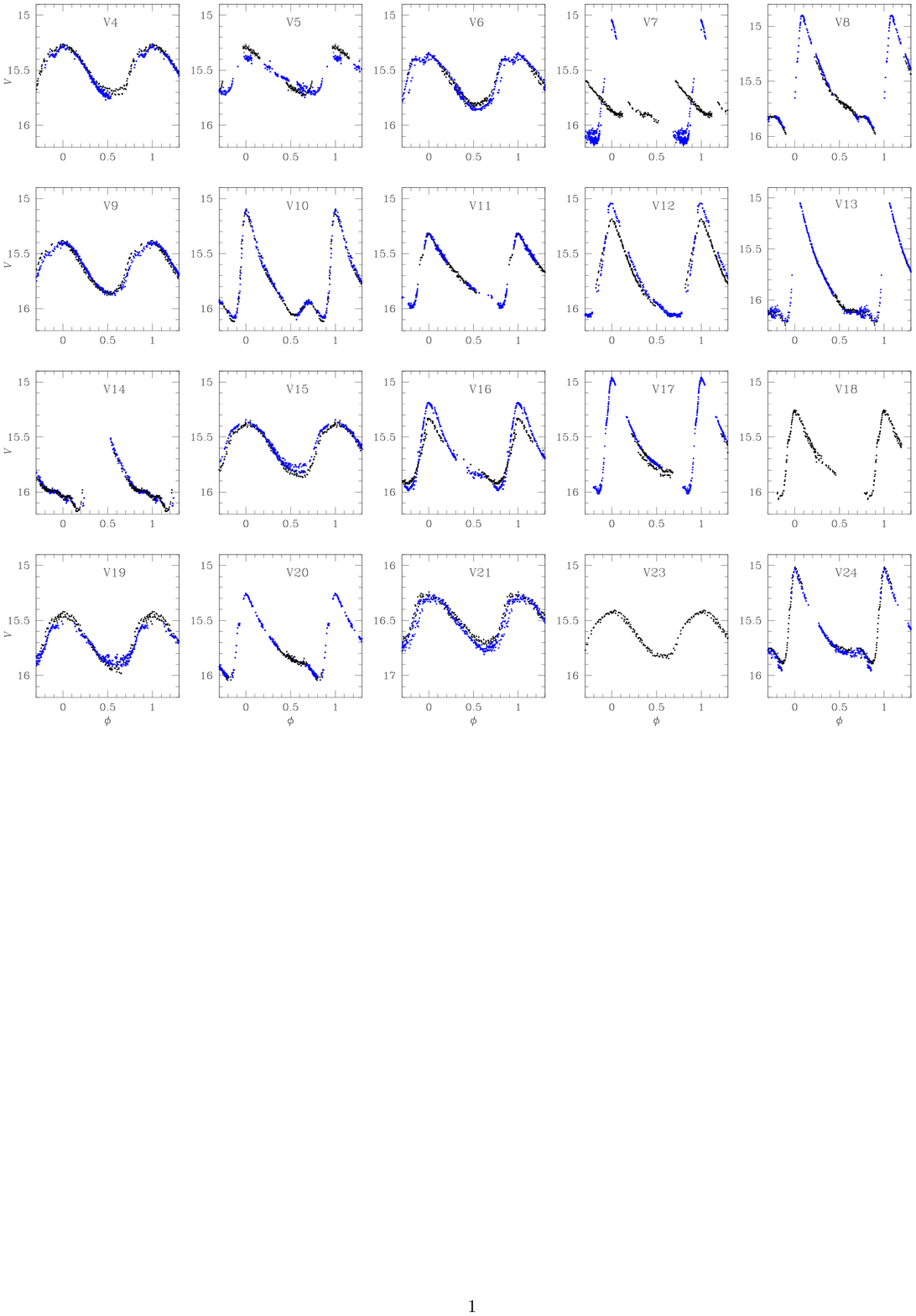}
\caption{Light curves of the 2015-2017 data,  except V23 for which the data
from Clement \&
Shelton (1997) were plotted since this star is not in the field of the 2015-2017 data.
Black symbols are used for the 2015-2016 data from Hanle. Blue symbols are for the
2017 data from SPM.
The light curves were phased with the new periods listed in Table \ref{variables}
(except for V5) and
the local epochs of the maximum light for a given data set as listed in Table
\ref{tab:maxima}, see $\S$ \ref{IND_STARS} for a discussion on individual stars.}

   \label{LCs}
\end{center}
\end{figure*}

In Fig. \ref{LCs} the $V$ light curves tabulated in Table \ref{PHOT15-17} and obtained
in 2015-2016 (black symbols) and 2017
(blue symbols) are displayed. They have provided a few recent times of maximum light
that extend the time-base to 82 years. They have been phased with corrected period
estimated below and the seasonal time of maximum. Note the consistency of the
2015-2016 and the 2017 data and the evident variation in amplitude in some stars,
likely due to the Blazhko effect. The employment of these light curves as indicators
of 
the mean metallicity and distance of the parental cluster, via the Fourier
decomposition, will be reported elsewhere. Comments on individual stars can be found
in $\S$ \ref{IND_STARS}.

\section{The $O-C$ approach to the secular period changes}
\label{OmC}

The observed minus the calculated ($O-C$) residuals of a given feature in the light
curve, as an indication of miscalculations or authentic variations of the
pulsation or orbital period, 
using a single given phase of the light curve as a reference, is a standard approach
that has been in use for many decades; for example, in Cepheids, RR Lyrae stars, and
contact
binary stars (e.g. Arellano Ferro et al. 1997; 2016; Coutts \& Sawyer Hogg 1969).
Then, it is convenient to select a feature that
facilitates the accurate determination of the phase. For RRLs, the maximum
brightness
is a good choice since it is well constrained, particularly for the RRab type, as 
opposed to the longer-lasting time of minimum. To predict the time of maximum let us
adopt an ephemeris of the
form

\begin{equation}
\label{ephem}
C = E_0 + P_0 N,
\end{equation}

\noindent
where $E_0$ is an adopted origin or epoch of reference, $P_0$ is the period at $E_0$,
and $N$ is
the
number of
cycles elapsed between $E_0$ and $C$. An initial estimate of the number of cycles,
between the observed time of maximum $O$ and the reference $E_0$ is simply

\begin{equation}
\label{cycles}
N = \left \lfloor \frac{O - E_0}{P_0} \right \rfloor,
\end{equation}

\noindent
where the incomplete brackets indicate the rounding down to the
nearest integer. However, it must be noted that if the time between $E_0$ and the
observed time of maximum $O$
is much larger than the period, and the period change rate is large enough, the $O-C$
difference can exceed one or more cycles, and there must be a correction for these
extra
cycles to obtain a correct $O-C$ diagram. This exercise may prove to be difficult 
if there are large gaps in the time-series, but it is rather straight forward
otherwise. 

\begin{table}
\footnotesize
\caption{Observed times of maximum light, $O$, for the RRLs in NGC 6171 and their
corresponding $O-C$ residuals calculated with the given ephemerides for each variable.
The sources of either the times of maximum light or the data employed to calculate
them, are coded in column 4 as follows: Oo (Oosterhoff 1938), CouSH
(Coutts \& Sawyer Hogg 1971), Man (Mannino 1961), Kuk (Kukarkin 1961), Di (Dickens
1970), Clem (Las Campanas, this paper),
ClSh (Clement \& Shelton 1997), Tab3 (Hanle and SPM, this paper).
This is an extract from the full table which is available in the electronic
version of the article (see Supporting Information).}
\centering
\begin{tabular}{lrrc}
\hline
Variable&$P_0 (days)$ & $E_0$ (HJD) & \\
\hline
V2& 0.571021&2442538.8397&\\
\hline
$O$ (HJD)&$O-C$ & No. of Cycles & source\\
   &  (days) &            &  \\
\hline
2427930.951& --0.0346&--25582.&Oo\\
2432004.682&   0.0340&--18448.&CouSH  \\
2437052.486&   0.0142&--9608.& Kuk \\
2437100.448&   0.0104&--9524.& Man \\
2439258.334&   0.0088&--5745.& Di  \\
2440693.874&   0.0025&--3231.& CouSH \\
2441448.747& --0.0140&--1909.& Clem \\ 
2442135.701&   0.0020&--706.&Clem \\
2442537.670& --0.0276&--2.&Clem \\
2442538.841&   0.0013& 0.&Clem \\
2442935.727&   0.0279&695.&Clem \\
2443273.736& --0.0075&1287.&Clem \\
2443632.899& --0.0165&1916.&Clem \\
2444020.635& --0.0037&2595.&Clem \\
2444371.805& --0.0115& 3210.&Clem \\
2444759.535& --0.0046&3889.&Clem \\
2445083.888&   0.0086&4457.&Clem \\
2445850.780&   0.0197&5800.&Clem \\
2446562.828&   0.0047&7047.&Clem \\
2447298.862&  -0.0071&8336.&Clem \\
\hline
\end{tabular}
\label{tab:maxima}
\end{table}

A plot of the number of cycles $N$ vs $O-C$, is usually referred as the $O-C$ 
diagram, and its
appearance can make evident secular variations of the period, or the fact
that the period $P_0$ used in the ephemerides is
wrong, in which case the distribution of the $O-C$ residuals is linear and tilted. For
NGC 6171, the distribution of the observations over the last 80 years enabled us to
estimate
accurately the number of cycles required in order to produce coherent $O-C$ diagrams.

Let us assume, as an initial model, a quadratic distribution of the $O-C$
residuals as a
function of time, represented by the number of cycles $N$ elapsed since the
initial epoch $E_0$. The linear and quadratic cases are then particular solutions
of the representation:

\begin{equation}
\label{parab}
O-C = A_0 + A_1 N + A_2 N^2,
\end{equation}

\noindent 
or,

\begin{equation}
\label{model}
O= (E_0 + A_0) + (P_0 + A_1) N + A_2 N^2.
\end{equation}

\begin{table*}
\caption{New periods and period change rates for RRLs in NGC~6171. 
The uncertainties in $\beta_0$ correspond to the uncertainty in the coefficient $A_2$
in eq. \ref{BETA_E}.}
\label{variables}
\centering
\begin{tabular}{cllllcc}
\hline
Variable& Variable & $P_0$& $E_0$ & $E_0 + A_0$& $P_0 + A_1$  &$\beta_0$\\

Star ID & Type     &    & (+2~400~000) &(+2~400~000) &   &$(O-C)$ \\
        &          & (days)   &   (HJD)      &(HJD)        & (days)& (d Myr$^{-1}$)\\
\hline
V2 & RRab & 0.5710   & 42538.841  & 42538.840  & 0.571021  &$-0.069\pm0.062$\\
V3 & RRab  & 0.5663   & 41844.561  & 41844.570  & 0.566344 &$+0.055\pm0.095$\\
V4 & RRc   & 0.282130 & 57528.3538 & 57528.3633 & 0.282132 &$+0.005\pm0.008$\\
V5 & RRab & 0.7024   & 57200.193 & 57200.3342 & 0.702376 &$-0.062\pm0.058$\\
V6 & RRc  & 0.259635 & 57527.2486 & 57527.2307 & 0.259627 &$+0.058\pm0.029$\\
V7 & RRab & 0.4975   & 44018.667  & 44018.635  & 0.497474 &$-0.155\pm0.117$\\
V8 & RRab & 0.5599   & 44371.599  & 44371.588  & 0.559922 &$+0.091\pm0.034$\\
V9 & RRc  & 0.3206   & 57200.2279 & 57200.2369 & 0.320601 &$+0.012\pm0.014$ \\
V10& RRab & 0.415506 & 57528.4208 & 57528.4707 & 0.4155586 & $\bf{-0.714 \pm 0.039}$\\
V11& RRab & 0.5928 & 57528.3744 & 57528.3821 & 0.592809 & $+0.096\pm0.022$\\
V12& RRab & 0.472833 & 49125.719  & 57527.331 & 0.472830 & $\bf{-0.756 \pm 0.168}$\\
V13& RRab & 0.4668 & 44371.865 & 44371.873 & 0.466797  & $+0.070\pm0.230$ \\
V14& RRab & 0.4816   & 43275.6195 & 43275.6356 &0.48162  &$+0.067\pm0.010$ \\
V15& RRc  & 0.288589 & 57200.1676 & 57200.0880 & 0.288590 &$+0.001\pm0.066$  \\
V16& RRab & 0.522798 & 57200.1894 & 57200.1503 & 0.522796 & $\bf{-1.088\pm 0.045}$\\
V17& RRab & 0.561168 & 41860.578 & 41860.593 & 0.5611675 & $\bf{+0.624 \pm0.021}$\\
V18& RRab & 0.561404 & 57528.4361 & 57528.4361 & 0.561404 & $+0.040\pm0.128$ \\
V19& RRc  & 0.278766 & 57528.3812 & 57528.3666 & 0.278762 & $+0.005\pm0.009$ \\
V20& RRab & 0.5781 & 41863.748 & 41863.765 & 0.578107&$-0.003\pm0.018$ \\
V21& RRc  & 0.258724 & 57201.2263 & 57201.203 & 0.258715 & $-0.004\pm0.053$\\
V23& RRc  & 0.323343 & 49477.617  & 49477.6189 & 0.323344 & $+0.076\pm0.056$\\ 
V24& RRab &0.523977  & 49123.856 & 49123.821 & 0.523947 & $-0.261\pm0.128$\\
\hline
\end{tabular}
\end{table*}

Taking the derivative, the period at any given $N$ is

\begin{equation}
\label{P-N}
P(N)=\frac{dO}{dN}= (P_0 + A_1) + 2A_2 N.
\end{equation}

From the above equations, it is straight forward to demonstrate that the period
change rate $\beta=\dot{P}$ at $N=0$ and $P=P_0$ is given by 

\begin{equation}
\label{BETA_E}
\beta = \beta_0 = \frac{2 A_2}{P_0},
\end{equation}

\noindent
and that if the $O-C$ distribution is linear, i.e. $A_2=0$, then the 
correct epoch and period are given by $E_0 + A_0$  and $P_0 + A_1$, respectively. A 
detailed derivation of the above equations can be found in Arellano Ferro
et al. (2016).

\section{Times of maximum brightness and the $O-C$ diagrams}
\label{Tmax}

\begin{figure*}
\begin{center}
\includegraphics[scale=0.95]{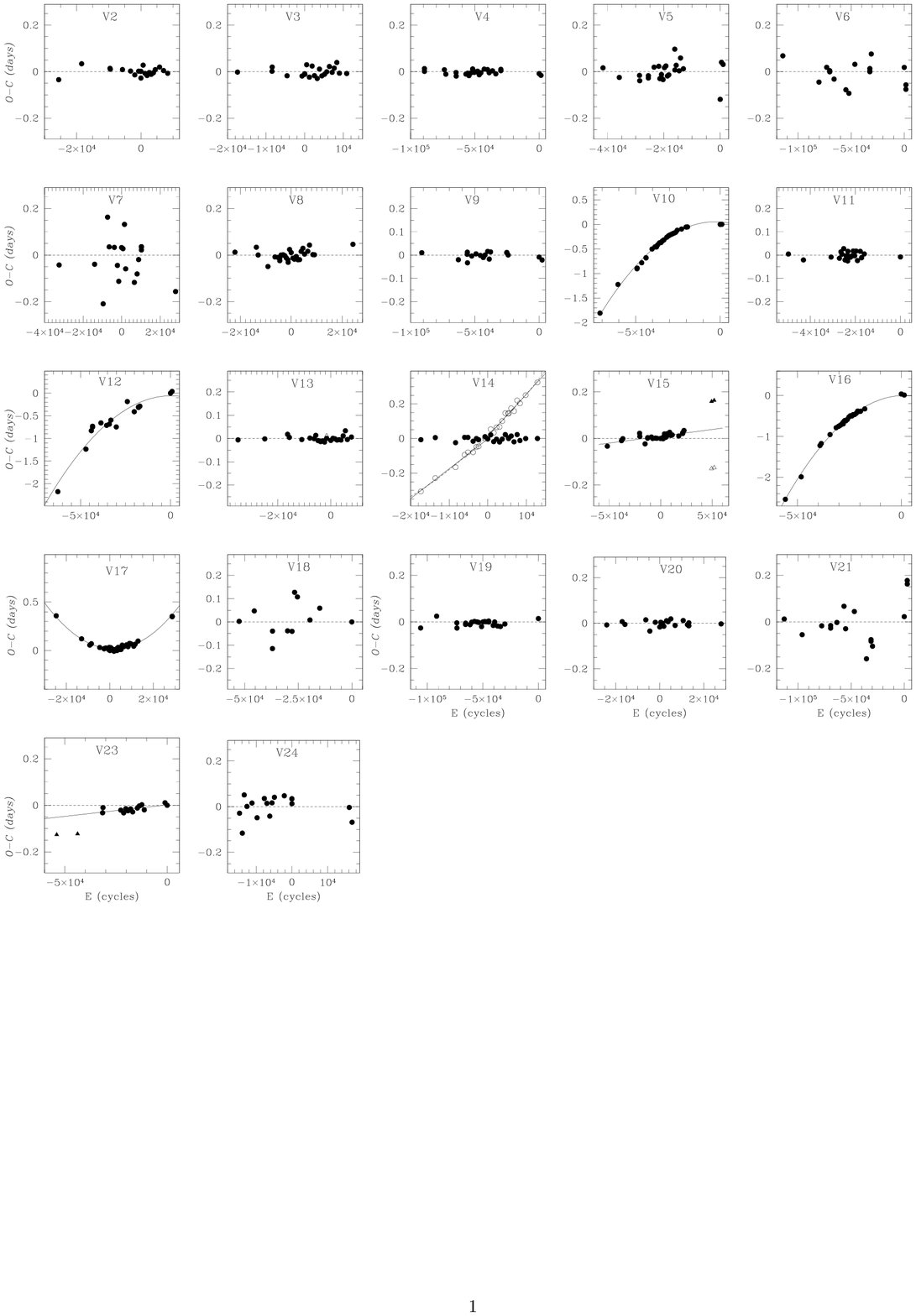}
\caption {$O-C$ residuals as a function of number of cycles (filled circles). For
stars with linear distributions, the $O-C$ residuals were calculated using the refined
ephemerides (columns 5 and 6, in Table \ref{variables}).
The $O-C$ residuals plotted with a triangle, filled and empty when two solutions were 
tried, stand out from the rest of the distribution and were not considered in the adopted
solutions. Solid lines represent the finally adopted period change solution for cases with 
$\beta$ different from zero. Dashed horizontal lines are plotted at ($O-C$)=0 as a
reference.
In the case of V5 we explicitly show the linear distribution showing the constancy of
its period. An apparent abrupt period change in V14 is illustrated by the empty
circles and two linear solutions. These and a few cases with peculiarities are
discussed in more detail in $\S$ \ref{IND_STARS}.}
\label{diagsOC}
\end{center}
\end{figure*}

We have estimated as many times of maximum brightness as possible with the available
data. When a light curve is covered near the maximum, estimating the time of the
maximum brightness is fairly straight forward and the uncertainty is small. Error bars
would be similar in size to the symbols in the $O-C$ diagrams. Only in a few cases,
interpolating between competing maxima was necessary. For the data sets covering
several years (e.g. Coutts \& Sawyer Hogg 1971 for 1946-1970 or data in Table 3 for
the years 1972-1991), we searched for clear times of maximum through the whole
collections and were able to recover numerous maxima, producing the highest density of
data in the $O-C$ diagrams in Fig. \ref{diagsOC}.

The complete collection of times of maximum light is given in Table \ref{tab:maxima}.
To calculate the $O-C$ residuals we proceeded as follows: first, an epoch with a well
covered light curve was identified and the period at that epoch was estimated. These
period and epoch were adopted as initial values $P_0$ and $E_0$. Generally, the data
from 2015 to 2016 were proper for this aim,
except for a few incomplete light curves particularly near the time of
maximum. In those cases, the data from Clement \& Shelton (1997) or from the
previously unpublished data from Clement (Table \ref {unpub}), were used.
For some cases with a linear distribution of $O-C$ values, we took the period from the
Catalogue of Variable Stars in Globular Clusters (CVSGC) (Clement et al.
2001; 2015 edition). While these periods are quoted to only four
digits and true periods may be slightly different,
the approach to the period correction is not sensitive to the selection of 
$P_0$ since a different $P_0$ will produce a different slope $A_1$ but the corrected
period $P_0 + A_1$ (see Eq. 5) will be the same. The adopted initial ephemerides are
summarised in
Table 4. The resulting $O-C$ diagrams are shown in Fig. \ref{diagsOC} for every
variable included in the present work. 

It is obvious from these diagrams, that there are only two types of $O-C$
distributions;
the linear distribution, which once the correct period is used in the time of maximum
predicting ephemerides (given in column 5 of Table \ref{variables}), produces a
horizontal distribution and implies a non-changing period, and the parabolic
distribution which implies a secular period variation whose rate can be calculated
with Eq. \ref{BETA_E}. It is worth noting at this point, that the $O-C$ approach to
the secular period changes is particularly sensitive to the counting of cycles, which
is very easy to lose when dealing with a long time-base and short-period stars, as is
the present case.
As the $O-C$ difference drifts, either because the assumed initial period is wrong or
because it is authentically changing, the calculated maximum may skip a few cycles
relative to the corresponding observed one. If this is not properly considered, the
$O-C$ diagram may show intriguing shapes which could be misinterpreted as irregular
period variations. This has already been stressed by Arellano Ferro et al. (2016) for
the case of the RRLs in M5. In the present case of NGC 6171, we do not see
any irregular variations but only constant periods (linear)
or a few parabolic secular variations (quadratic). It is pointed out however, that
there
are cases where the horizontal $O-C$ distributions display a considerable scatter
(e.g. V6, V7, V18, V21). This may be the consequence of stochastic fluctuations of
the period and/or of uncertainties in the estimations of the times of maximum light
due to the limited quality of the observations. 

Table \ref{variables} summarises the initial assumed ephemerides, the corrected
periods
and epochs and, in the corresponding cases, the calculated period change rate
$\beta$.

\subsection{Consistency test for the refined periods and period change rates}
\label{phasing}

Once a period change rate has been calculated (quadratic case), or the period has
been duly corrected (linear case), a natural test is to phase
the light curve with the new ephemeris. In Fig. \ref{LCs}, the light curves
from 2015-2016 are phased either with corrected period (linear cases) or with
$P(N)$ for the corresponding $N = 0$ in Eq. \ref{P-N} (parabolic cases).

In fact, for a secularly changing period, Eq. \ref{P-N} allows to calculate the period
at any given number of cycles $N$ elapsed from the origin $E_0$. Each corresponding
period should properly phase the data taken
at that vlue of $N$. As an example, in Fig. \ref{V12} we phase the light
curve of star V12 over the last 82 years, using the "local" periods and epochs as
predicted by the
parabola in Fig. \ref{diagsOC} and Eq. \ref{P-N}, and listed in Table \ref{V12sec}. In
all cases, the light curve and time of maximum are consistent with the secular period
change. The appearance of the curve is only limited by the quality of the 
photometry of a given data set.

\subsection{Comments on individual stars}
\label{IND_STARS}

While generally the $O-C$ diagrams in Fig.\ref{diagsOC} show a clear period behavior, 
in a few cases the $O-C$ distribution may admit alternative solutions, as we comment
in the following paragraphs.
Since the previous study of the period changes in NGC 6171 by Coutts \& Sawyer
Hogg (1971), good quality data have been obtained, and hence richer $O-C$ diagrams
can be produced, a detailed comparison with the results  of these authors is
probably inadequate. Some comments on specific stars might however be in order.

V2 and V3. These RRLs are not included in the field of view of our images and the
historical data are scarce. However, the data in Table \ref{unpub} enable the
estimation of numerous times of maximum light and hence the analysis of the secular
behaviour of the period.

V5. The $O-C$ residuals in Fig. \ref{diagsOC} display a clear linear
distribution, leading to a refined period of 0.702376 d. A negative period
change rate and an abrupt negative period change have been reported for this star by
Coutts \& Sawyer Hogg (1971) and Gryzunova (1979a) respectively. Our solution displays
a rather constant period.
The estimated maximum from 2015 data shows a significant and unexpected phase
displacement and was
not considered in the adopted solution. We noted however that the predicted period in
Table 5 (0.702376 d) fails to phase properly the light curves from 2015-2017,
for
which a shorter period, 0.695248 d, had to be invoked to scale the curves well. We
do not have a clear explanation for this behaviour and speculate that the star
might have undergone a stochastic variation of its period. The amplitude variation
between 2015-16 and 2017 should be noted.

V6. Although the $O-C$ distributions of this star suggest a linear
distribution, the scatter is significant and is probably due to the bump near maximum
which makes the estimation of the time of maximum brightness inaccurate.

V7. In our data the light curve of this star displays a large difference between
2015-2016 and 2017 and suggests a large amplitude modulation as observed in stars
with the Blazhko effect. Clement \& Shelton (1997) noticed the cycle to cycle
variations near minimum light,
and the peculiar harmonics amplitude rations relative to other RRab stars in the
cluster. Stars with Blazhko effect often display not only amplitude but also phase
modulations, which likely explains the large scatter in the $O-C$ diagram for this
star.

V10. A period increase was reported by Coutts \& Sawyer Hogg (1971) who 
calculated $\beta = 1.1$ d~Myr$^{-1}$. The diagram in Fig. \ref{diagsOC} shows a  
decreasing nature of the period at a rate $\beta = -0.714 \pm
0.039$ d~Myr$^{-1}$. The discrepancy between the two investigations is caused by an
error in
the period due to an uncertainty in the number of cycle counts in the 
CSH study. The richer  $O-C$ diagram in our current study has resolved this ambiguity.

\begin{figure*}
\begin{center}
\includegraphics[scale=1.]{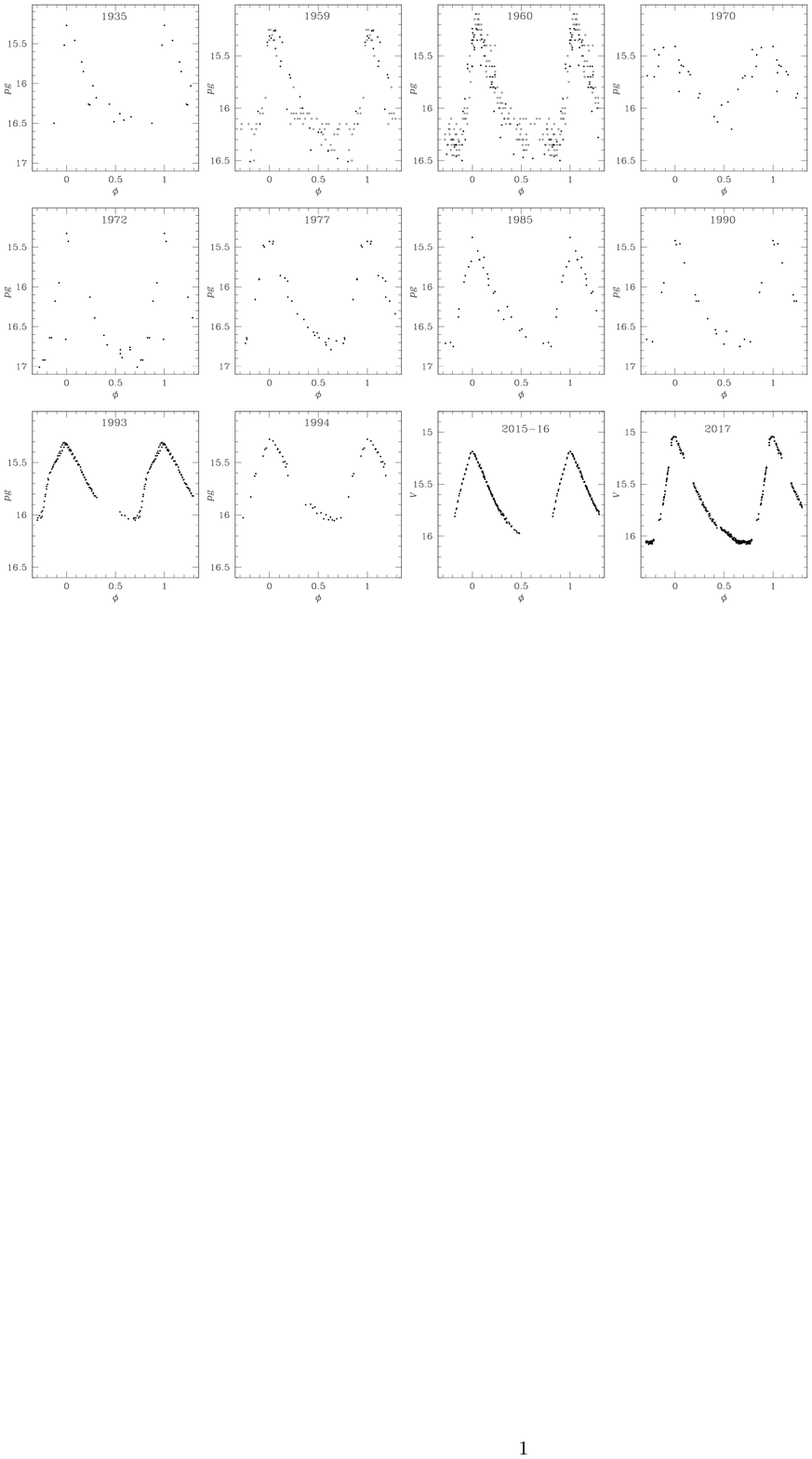}
\caption{A selection of light curves of V12 over the past 82 years phased with the
corresponding
period and epoch as predicted by the parabola in Fig \ref{diagsOC} and Eq. \ref{P-N}.
Note the variations in amplitude.
These ephemerides are listed in Table \ref{V12sec}.}
\label{V12}
\end{center}
\end{figure*}

V12. Our analysis of this star shows a decreasing period with 
$\beta = -0.756 \pm 0.168$ d~Myr$^{-1}$. The $O-C$ diagram however displays a
significant dispersion. We note that Clement \& Shelton
(1997) found the Fourier parameters, particularly $\phi_3$ and $\phi_4$,
to be peculiar among those in other RRab stars. A close inspection of Fig. \ref{V12}
shows a distinctive slope change on the rising branch in 1993 which is not apparent in
the other light curves. This suggests that the light curve shape might undergo
secular variations and probably stochastic oscillations of the time of maximum. It
should be noted the amplitude variation between 2015-2016 and 2017 data, confirming
the amplitude modulations reported by Clement \& Shelton (1997).

\begin{table}
\caption{A selection of epochs and periods of star V12 over 80 years. The phased
light curves are shown in Fig. 3.}
\label{V12sec}
\centering
\begin{tabular}{ccc}
\hline
E0  &  P (days) & year \\
\hline
2427931.957  &  0.47289860 & 1935\\
2436728.475  &  0.47287822 & 1959\\
2437050.549  &  0.47287747 & 1960\\
2440747.669  &  0.47286889 & 1970\\
2441454.580  &  0.4728665 & 1972\\
2443281.546  &  0.4728627 & 1977\\
2446185.774  &  0.4728567 & 1985\\
2448011.631  &  0.4728529 & 1990\\
2449125.719  &  0.47284945 & 1993\\
2449482.737  &  0.47284862 & 1994\\
2457527.331  &  0.47282997 & 2015-2016\\
2457939.681  &  0.47283244 & 2017\\
\hline
\end{tabular}
\end{table}

V14. The $O-C$ residuals for this star show a peculiar change in slope at about
HJD 2443275.6 d or May 1977 if a period of 0.4816 d is used. Although this slight
change of slope could also be
fitted by a parabola, implying $\beta = +0.32 \pm 0.07$ d~Myr$^{-1}$, we rather prefer
the
slight period change and hence the two slopes depicted. The two slopes would
imply corrected periods $P1=0.481617$ d and $P1=0.481624$ d which are identical to
the fifth digit. We failed observing the star near maximum brightness between 2015
and 2017.

V15. A slightly tilted $O-C$ diagram is found with two discrepant values in 2015 and
2017, which are otherwise consistent among them. This may suggest an abrupt period
change which is to be confirmed in the future if new times of maximum light become
available.

V16. This is a clear and strongly period decreasing star with $\beta = -1.088 \pm
0.045$. Coutts \& Sawyer Hogg (1971) also found a similar result and calculated a
period decrease rate of -1.6 d~Myr$^{-1}$.

V17. The $O-C$ diagram in Fig. \ref{diagsOC} shows a positive parabola and a
corresponding $\beta = +0.748 \pm 0.021$. This is the only star in the sample for
which
we have found a positive period increase that, if ascribed to evolution, may indicate
a red ward evolution.

V21. In spite of the fact that this is a field star (Dickens 1970, Cudworth et al.
1992)
projected on the cluster field, it was included in the present due to the fact that
similar to the cluster members, photometric coverage was available for 82 years.
Hence, although the scatter in the $O-C$ diagram is substantial, its period has been
refined and reported in Table \ref{variables}.

V22. This star is not included in the present study. V22 is
not a cluster member (Sawyer Hogg 1973) and out of the field of most studies, except
of that of Oosterhoff (1938). Hence the historical data are very scarce.

V23. The $O-C$ diagram shows a linear distribution with a small tilt which implies
a tiny correction to the period. The 1935 light curve from Oosterhoff (1938)
is very scattered and we could not estimate a reliable time of maximum. We note
that the two oldest maxima, corresponding to
data from Coutts \& Sawyer Hogg (1971) from July 1946 and July 1955, are not fitted by
the more recent linear distribution, however these are two {\it bona-fide} maxima from
a rather scattered light curve. We note that in the paper by Dickens (1970) the star
labeled as V1 in fact corresponds to V23.

\section{Period changes and evolution in the HB}
\label{EVOL_HB}

As stars evolve across the instability strip (IS), their pulsation
period should either increase or decrease if evolution
is towards the red or the blue, respectively.
However, other non-evolutive reasons for period changes have been suggested,
such as stochastic variations (Balazs-Detre \& Detre 1965) or mixing 
events in the core of a star at the HB that may alter the hydrostatic structure and
pulsation period (Sweigart \& Renzini 1979). 
Also, irregular and complicated secular period variations have been claimed for some
RR Lyrae stars (e.g. Szeidl
et al. 2011, Jurcsik et al. 2001), which indeed would be difficult to reconcile
with stellar evolution exclusively. However, for the RR Lyrae stars in M5, it has
been argued that
there is no need to claim for irregular period variations since an
improper counting of cycles,
particularly in long time baseline sets of times of maximum light, may be responsible
for apparent irregularities (Arellano Ferro et al. 2016).

At present, there is a large evidence that there is no preferential positive or
negative 
values of $\beta$ in the RRLs in a given cluster, and for a summary the
reader is referred to the discussion of Arellano Ferro et al. (2016). Perhaps, the 
exceptions are $\omega$ Cen for which an average of $\beta
= +0.170\pm0.561$ d~Myr$^{-1}$ can be calculated from table 6 of Jurcsik et al.
(2001), and
IC~4499
with $\beta= +0.29 \pm 0.60$ d~Myr$^{-1}$ from Kunder et al (2011) (their Table 1
without
three extreme cases). Also, in the extensive investigations on secular period
variations in globular clusters (e.g. Silbermann \& Smith 1995; Corwin \& Carney
2001; Arellano Ferro et al. 2016) no significant differences have been found for the
average values of $\beta$ for the populations of RRab and RRc stars .

Models of the HB calculated by Lee (1991) and Catelan et al. (2004) confirm that
positive period change rates of evolutionary origin
occur mostly in globular clusters with blue HB structures, i.e. with large values of the HB 
structure parameter $\cal L$ $= (B-R)/(B+V+R)$, where $B, V$ and $R$ are the number of stars to the blue,
inside and to the red of the instability strip respectively. Figure 15 of Catelan (2009) displays 
such behaviour of $\beta$ as a function of $\cal L$ and shows that in red HB clusters the average
value of $\beta$ should be about zero. NGC 6171 has a very red HB, with $\cal L$ $= -0.74$,
hence the above models predict an average $\beta \sim 0$ d Myr$^{-1}$. In fact, the overall average of $\beta$ values
in Table \ref{variables} is $-0.086\pm 0.346$ d Myr$^{-1}$, which given the typical uncertainties of $\beta$ is not
significantly different from zero.

Even in clusters like M3 and M5 where the overall period change rates average nearly
zero, as
predicted by canonical models, it has been common to isolate individual cases with
significantly large values of $\beta$. 
In the case of NGC 6171, we found large secular period change rates in 4 stars in a sample of 22.
Three of these have large negative values of $\beta$, which implies
evolution to the blue. The only period increasing case is V17 with $\beta =
+0.624 \pm 0.021$. This period rate is comparable to the rate found in some
RRLs in other clusters; in M5 ($\cal L$=+0.31) for example, V8 (+0.474 d Myr$^{-1}$), V7 (+0.474
d Myr$^{-1}$) and V25 (+0.933 d Myr$^{-1}$), or the more moderate V77 (+0.340 d
Myr$^{-1}$ ), V87 (+0.369 d Myr$^{-1}$) and V90 (+0.114  d Myr$^{-1}$) for which,
arguments in favour of they being stars in a truly advanced evolution have been
offered
(Arellano Ferro et al. 2016); in M3 ($\cal L$=+0.18) we have V10 (0.385 d Myr$^{-1}$), V47 (+0.393d
Myr$^{-1}$), V69 (+0.414 d Myr$^{-1}$), V83 (+0.345 d Myr$^{-1}$) (Corwin \& Carney
2001).
 Jurcsik et al. (2001) calculated period change rates across the IS between
$-0.026$ and +0.745 d Myr$^{-1}$ based on post-HB evolutionary tracks of Dorman
(1992) for [Fe/H] = $-1.48$ and masses of 0.60 to 0.66 M$_{\odot}$. Thus, all the
above
quoted positive period changes may be consistent with an evolutionary origin, also for
our present case of V17. 

On the other hand, as discussed by Silva-Aguirre et al. (2008), pre-ZAHB
stars crossing the instability strip at high evolving rates may have values of 
$\beta \sim -0.3$ d Myr$^{-1}$ but can reach 
values $< -0.8 $ d~Myr$^{-1}$. From the calculations of Jurcsik et al. (2001)
based on Dorman's (1992) post-ZAHB models, the fastest blueward evolution reaches the rate of
$-0.7\times10 ^{-10}$ d d$^{-1}$ or about $-0.026$ d Myr$^{-1}$. Thus, it is tempting to suggest
that variables V10, V12 and V16 in NGC 6171 are examples of pre-ZAHB stars. We should 
consider however that, according to the statistics produced by the simulations of Silva-Aguirre et al. (2008)
for the case of M3, only 1 pre-ZAHB is expected every 60 {\it bona-fide} HB stars, 
and only 22\% of them would fall in the instability strip, where they can be disguised
as RRLs. Assuming that
these statistics hold for NGC 6171, with about 110 stars in the HB, it implies that not even one 
pre-ZAHB RR Lyrae-like pulsator should be found. Nevertheless,
RR Lyrae with large negative values of $\beta$ is a rather common feature in several clusters. 
In M3 itself there are 5 stars with $\beta < -0.4$ d~Myr$^{-1}$ (Corwin \& Carney 2001), in M5 there are 
5 stars with  $\beta < -0.3$ d~Myr$^{-1}$ (Arellano Ferro et al. 2016) and 4 in NGC 6934 (Stagg \& Wehlau
1980). Thus, it is probably not unlikely that V10, V12 and V16 in NGC 6171 are indeed
pre-ZAHB stars.

Perhaps the most remarkable result in the present paper is the high percentage of
stars with stable period. It should be noticed that the four stars with changing
period in NGC~6171
are RRab stars, and that 18 of the 22 stars studied, i.e. 82\%,
have retained a constant period for at least 80 years. 
In M5, 34\% of its RRLs were found to have unchanging periods over a 100 year interval
(Arellano Ferro et al. 2016).
Naturally, the question of whether this result would be subject to change after one or
two decades of accurate estimation of times of maximum may be posed. Note that the
uncertainties in the beta values in Table \ref{variables} are generally of a few hundreds 
of d~Myr$^{-1}$, and that, if the linear distributions in Fig. \ref{diagsOC} are forcibly
fitted with a parabola, the quadratic term leads to very small and non-significant values of
$\beta$. We conclude then, that with the data on
hand, we are unable to detect period variations, positive or negative, below these
limits. Thus, period changes for stars evolving very near the HB at very low rates
may pass undetected.

\section{Summary}
\label{summary}

Pulsation period changes have been analysed via the times of maximum light for 22 RRLs
in NGC 6171. Archival data collected from the literature, previously unpublished
data spanning 19 years, and recent CCD observations enable a span of up
to 82 years for most of the sample stars, which makes this work the first
significant study of period changes in NGC 6171. 
Secular period variations were found for 4 stars, three with significant decreasing
periods and one (V17) with increasing period. No signs of irregular period variations
were found in the RRLs of this cluster but
instead they all have either a remarkably stable period or a secular
period change that can be represented by a parabolic $O-C$ diagram. 

The overall average of the period change rates found in NGC 6171 is not
significantly 
different from zero, as expected from the canonical evolutionary models of the HB 
for a cluster with a red HB. Not withstanding this fact, individual stars with large
positive and negative period changes have been found, a trend also observed in M3 and
M5.

In NGC 6171, we have found a single case with positive $\beta$ (V17) which
seems to be consistent with the period change rate expected in a truly
advanced stage of evolution towards the AGB. On the
contrary, a few cases emerged with values of
$\beta$ significantly negative, which cannot be reconciled with
post-HB evolutionary
predictions, and that may be examples of pre-core-helium-burning stars on their
contraction towards the ZAHB.
The majority of the RRLs in this cluster, both RRab and RRc, display a stable
period for at least 82 years well within the uncertainties of the $O-C$ approach.
Under the paradigm that period changes are a consequence
of stellar evolution, it must be concluded that these stable stars are evolving
very slowly and their putative period changes are, given the data presently
available, undetectable by the approach described in this work.

\section*{Acknowledgments}
We are grateful to Prof. Christine M. Clement for encouraging this work and for
supplying us with her collection of relevant historical data of NGC 6171, and her own
extensive unpublished observations taken between 1972 and 1991 and allowing us to
publish them in the present paper. Her comments and suggestions to the manuscript are
gratefully appreciated. We are also indebted with the anonymous referee for his/her
useful suggestions and enlightening comments.
We acknowledge the financial support from  DGAPA-UNAM, M\'exico via grants 
IN106615-17, IN105115 and from CONACyT (M\'exico). PR is grateful for the financial
support from
the PREI program of the National University of M\'exico and the hospitality of
the Instituto de Astronom\'ia (UNAM).  PR warmly acknowledges the financial support 
of the CDCHTA - Universidad de Los Andes (ULA) through project C-1992-17-05-B. We are
thankful to
Carlos Chavarr\'ia for his help with some of the obervations in SPM.
We have made an extensive use of the SIMBAD and ADS services, for which we
are thankful.

\end{document}